\documentstyle[12pt,aasms4]{article}
\begin{document}
\title{QSOs in the combined SDSS/GALEX database}
\author{J.B.Hutchings }
\affil{Herzberg Institute of Astrophysics, 5071 West Saanich Rd.
Victoria, B.C. V9E 2E7, Canada; john.hutchings@nrc.ca}
\author{L. Bianchi}
\affil{Dept of Physics and Astronomy, Johns Hopkins University, 
3400 N. Charles St., Baltimore, MD21218}
\begin{abstract}
We discuss selection of QSO candidates from the combined SDSS and GALEX catalogues.
We discuss properties of QSOs within the combined sample, and note 
uncertainties in number counts and completeness, compared with other
SDSS-based samples. We discuss colour and other properties with
redshift within the sample and the SEDs for subsets. We estimate the 
numbers of faint QSOs that are classified as extended objects in the 
SDSS, and consequent uncertainties that follow.

\end{abstract}

\section{Introduction}

   The SDSS has produced major catalogues of optically selected objects
which are of interest for many studies. The SDSS has also produced the
largest collection of QSOs to date, by a large margin. The next step is
to identify and verify which they are. While some of
these objects have SDSS spectra which can identify them as QSOs, 
they are a small fraction compared with the GALEX-SDSS QSO candidate
selection of Bianchi et al. (2007), 
and it is of interest to estimate the space density,
number counts with magnitude, and luminosity distribution of QSOs, based
on this larger and fainter dataset.
It is also interesting to examine the parameter space
favoured by the different selection criteria. 

    Like all optically selected QSO surveys, there are selection effects 
that must be understood, in order to produce these distributions correctly.
The SDSS filters alone do not allow a clean separation of QSOs from other
blue objects. The GALEX UV surveys add very useful wavelength leverage
to isolating QSOs, when combined with optical data (see Figs 5-7 of 
Bianchi et al 2007). In this paper, we discuss
QSO candidates based on the combined SDSS and GALEX survey databases. 
Specifically, we use the catalog of low-redshift QSO candidates selected 
from a matched source catalog similar to that constructed by Bianchi 
et al. (2007) from the GALEX GR1 and SDSS DR3 data releases. The overlap 
areas are 363 square degrees in the GALEX AIS 
(All-sky Imaging Survey) and 83 square degrees in the GALEX MIS (Medium Imaging Survey). 
The AIS has typical exposures of about 100 sec and reaches objects of
limiting (5 $\sigma$) NUV-band AB magnitudes of about 20.8 
while the MIS has average exposures 
of 1500 secs, reaching objects with NUV magnitudes of about 22.7. 

   Based on initial colour cuts in FUV-NUV, NUV-r as in Bianchi et al (2007),
and photometric error cuts of 0.3 mag in FUV, NUV, and r,
we begin with about 34000 objects from the MIS and 6000 from the AIS.
Of these, 17\% and 55\% respectively are classified as point sources.
A further cut in u magnitude at 24.2 (or u band error 1.5 magnitudes)
could eliminate a few more  sources that are probably 
spurious in that band, but this amounts to less than 4\% of the MIS 
point sources and none at all of the other samples. In the sections below
we describe further cuts to these samples, to match the properties of
the subsets of about 800 in the MIS and 1600 in the AIS that are known QSOs,
with redshifts, from  SDSS spectra. We then discuss the properties of the
resulting catalogues in terms of the number counts, space density, and
luminosity function of QSOs over the redshift range about 0.1 to 2.
Table 1 gives a summary of the catalogues we discuss.

 The Bianchi et al (2007) selection of QSOs
should include essentially all QSO between z=0.5-1.5 (which are
also the least contaminated by foreground stars), but it's definitely
incomplete for z around 0 and $>$1.6, because the natural spread of SEDs
around an average template, and extinction effects, blur QSO and stellar
loci together.
In other words, QSOs around z=0  and $>$1.5 share the stellar locus in
part, so their colour selection was made such as to minimize the stellar
contamination for z=0, at the expense of missing a fraction of QSOs.
In this paper we discuss ways to get more complete QSO candidate lists
and some values for their magnitude and redshift distribution.

\section{Properties of known QSOs}

   We use the photometric properties of the 
spectroscopically confirmed QSOs included in our candidate QSO catalogues to 
estimate the QSO population and properties of the full GALEX+SDSS
catalogues. Figure 1 shows some of the photometric properties of the 
identified QSOs with redshift.

We recall that these QSO candidates were selected by Bianchi
et al. (2007) from the NUV-r, FUV-NUV color-color diagram (see
their Figure 7, bottom panels), where the colours of QSOs based
on templates from previous known samples separate from most types of
stellar objects and galaxies. Therefore, this work addresses the
selection of QSOs with similar properties.  In future work we
will investigate the possible existence and characteristics of 
QSO samples with differing SEDs in the UV.

At redshifts 0.5 and higher, there is a very tight and systematic
change of g-r colour with redshift. Lower redshift QSOs show scatter to
larger values, but a g-r range of -0.2 to 0.4 includes over 97\% of them.
There is a similar tight relationship with r-i. Figure 1 shows these
samples. 

   The lower panel of Fig 1 shows the correlation involving UV
flux and redshift.
The quantity plotted is the following combination of magnitudes:
4r+2g-2i-z-3FUV. This has a tighter and more single-valued dependence on
redshift than any simple difference of two magnitudes, such as FUV-u or 
NUV-r, etc. However, these other combinations
of SDSS and GALEX magnitudes yield similar trends with redshift. In general, 
the photometric errrors on u are larger than for r, so we use the boundary 
of NUV-r as a further cut on the large photometric catalogues. The lower
limit of NUV-r is 0 for the MIS spectroscopic sample and -0.3 for the 
AIS sample. The 
u-band errors in the identified QSO subset are all small, so we may use 
this index as a redshift estimator in the photometric catalogues, provided we 
also eliminate objects with u-band errors of 0.5mag or larger. Table 1 
shows the dataset sizes with these cuts.

\section{Photometric selection in redshift bins}

  From Fig 1, using cuts at g-r=0.15 or r-i=0.1, we can separate the QSOs into
redshift bins roughly 0.3 to 0.7, 0.9 to 1.6, and above 1.5. We restrict 
the overall ranges in these colours to -0.15 to 0.4 as almost all 
identified QSOs lie in this range. 

  The cleanest redshift cut is the bin 0.9 to 1.6, using g-r. There is contamination by a few low redshift QSOs, and these can be eliminated by 
cutting the red end of the UV-optical index. This process includes 93\% 
of the identified QSOs in this redshift range. Applying the same
cuts to the entire MIS point source (MISP) sample, we find a total of 
1646 candidates, including 412 of known redshift. For the AIS point source
(AISP) sample the cuts retain 82\% of the known QSOs sample in the redshift
range. Applying the same cuts to the entire AISP sample, yields an extra 
153 candidates, which scales to
187 allowing for this incompleteness. Since the total number is 623, the
identification of QSOs in the AISP sample appears to be complete at the 75\%
level.

The redshift range 0.9 to 1.6 is
under-represented in many samples because of the ground based bandpasses
and wavelengths of key emission lines. If we can separate the lower 
redshift QSOs (0.2 to 0.7) which have the same colours, we can derive
new values for number counts and magnitudes for these redshift bins.
The g-r and r-i cuts on the known QSOs yield redshift bins above
1.5 and below 0.7, roughly. We can use the UV-visible colour index
to separate them, but there is enough scatter that we inevitably have some
overlap. Thus, the high redshift bin loses some candidates and accretes
some low redshift objects. The optimum cut for high redshift QSOs yields 
close to the correct total numbers, but with some 10-20\% moved into and out
of the redshift bin of interest. We looked at the mean magnitudes and
magnitude distributions of the low and high redshift QSOs from the 
spectroscopic catalogue, and find no significant differences (while the
intermediate redshift objects do have a different distribution). Thus
we consider that we can get lower limits to number counts but good magnitude
distributions for low and high redshift candidates, although there will
be some cross-contamination in the lists. 

   For the redshift greater than 1.5 objects, the MISP has 1120 objects in
the candidates list, if the contamination average is the same. Of these,
200 have known redshifts. For the AISP sample, the numbers are 230 total
candidates, of which 200 are already known. Thus, again we find almost
complete identification of the QSOs in the AISP sample, in the SDSS
spectroscopic database. This is mainly a reflection  of the deeper
limits in the MIS sample compared with the SDSS spectroscopic database.
Figure 2 shows histograms of the magnitude distributions of these QSO
candidates, compared with those for the identified QSOs, and the 
completeness limits of the spectroscopic data are evident. The fact
that fewer faint candidates exist for the low and high redshift group
(top panels) presumably is because there are few faint low redshift objects,
and the incompleteness of both the low and the high redshift sample.

Table 2 shows some mean magnitude and colour indices for the different
samples we discuss.  The brighter limit of the AIS compared with the MIS is
evident - also the difference between the spectroscopic and MIS limits.
The colour difference between the candidates and spectroscopically
confirmed QSOs are less obvious, and may indicate some differences in 
populations with redshift, as well as magnitude limits and errors.

 Among the sample of extended objects with spectra, we do not expect to 
find many QSOs, since they are normally registered as point sources. 
Indeed, there are only 34 (of 29287) spectroscopically confirmed QSOs 
in the MISE spectroscopic database, and 109 of 2633 in the AISE. The same 
colour cuts applied to these samples, reduce the QSOs counts to only 3, 
so this is clearly not going to produce any significant number of QSOs,
at least to the magnitude limit of the SDSS spectra. This may not be true
of the fainter sources, as we discuss in the next paragraph. The bright 
ones that are found have red colours and low redshift, indicating obscured
nuclei, so that the host galaxies are more likely to be seen as resolved
in the SDSS.  While red QSOs may be a significant population (e.g. Hutchings 
et al 2006), this UV-optical database is not effective at finding them.

If we apply the r-i and g-r with NUV-r colour cuts to the MISE sample we
find the total `candidates' shown at the bottom of Table 1. These are much
larger numbers, so clearly they represent more than just QSOs. 
Further cuts to the r-i and g-r range correspond to redshifts 0.6 to 1.0:
we get slightly fewer than the full range g-r cut - some 4100 candidates.
These numbers embody reasonable cuts to the g, i, and u band errors,
but are very sensitive to the error values for u band. Overall, we feel these
numbers suggest that of order 10\% to 20\% of the MISE sample may be faint QSOs.
The mean g magnitudes of these samples are some 0.5 magnitudes fainter
than the MISP candidates (Table 2). 

\section{Space density and completeness}

   Figure 3 shows the number counts per square degree of sky for the
QSO candidates we have discussed. It is notable that the counts do not
match for the AIS and MIS, at the bright end of the distributions. This
indicates that there is incompleteness in the AIS list, presumably because
of rejection by error bars in the weakest signal bands - u and z. The
error bars for the MIS sample are much smaller because of their 15 times
larger exposure times. In addition, the GALEX magnitude errors are
larger for the AIS sample than the MIS, so there may be some systematic
difference in the UV depth too. Overall, we thus consider that the
MIS sample is complete to the maximum bin magnitudes (g=21), and it falls
off fast fainter than that. This diagram may be compared with Figure 11
of Bianchi et al, for their broader selection of QSOs. 

   Figure 4 shows the median magnitudes from each of the 6 bands, for
the various QSO samples, together with the magnitude differences between
them. We have compared a bright subset of the MIS samples, with g magnitudes
less than 19, for direct comparison with the AIS samples, which have very
similar mean brightness. The two subsets are divided into redshift
bins as Figure 3. In both sets, the bright MIS sample is redder than the
AIS - i.e. it reaches fainter FUV magnitudes, while the r to z bands
are brighter. This difference is more marked in the intermediate redshift 
set. The SED for the MIS total samples are essentially the same as the
bright subsamples. This indicates that the longer exposures of the MIS
gains more sensitivity with respect to the AIS progressively as we go to
shorter wavelengths.  The colour difference plots in the lower panels
show the same things: the difference between MIS and AIS increases
as we go to shorter wevelengths, or the MIS is more sensitive to blue
objects. This is consistent with the relative incompleteness of the
AIS sample in Figure 3.

   Note that the presence of strong emission at Ly$\alpha$ and C IV
would make the median SED brighter in the NUV and u bands for the 
redshift 0.8 to 1.8 sample and the u band for the higher redshift sample.
The Lyman edge drop would affect the FUV and NUV bands in the opposite 
sense. Allowing for these effects only increases the difference in MIS 
SED between the two redshift bins. The higher redshift QSOs are bluer, 
as may be expected for UV-bright rest frame SEDs.  

\section{Discussion}

   The space density of QSOs with the SDSS has been discussed by a number
of authors (Richards et al 2004, 2005, 2006, vanden Berk et al 2005). The
UV properties and UV-selection have been discussed by Bianchi et al 2006, 
and Trammel et al 2006.  We have compared our candidate counts with all these
and find several points of interest, as they do not agree well. Figure 5 
shows some comparisons. The MIS sample we have is different from the
large sample of Richards et al, and their subsequent ones, in having 
lower space density peak, but more in the bright end of the distributions.

We have noted that our AIS sample gives lower source counts than the MIS.
Looking at the magnitude 17-18 range, where our sample has an excess compared
with the Richards et al candidates, we have a very high fraction of spectroscopically
confirmed QSOs, so the difference seems robust. It is possible that
some of these are low luminosity QSOs or Seyferts, somehow excluded by 
the Richards et al selection. At the fainter end, if we assume that some
of the MISE sample are misidentified QSO point sources, we boost mainly the faint end of the distribution, as discussed in the previous section. 

One issue is that of the true sky coverage of the various samples discussed.
It may be that they have been mis-estimated in some cases. 
Figure 5 shows our numbers from a small subset taken from the Richards et al
catalogue (about 65 sq degrees, with some 3700 objects). It also shows the 
published values for the whole Richards sample, which suggests that their 
sky area is
about right. Our subsample is similar to their total sample plots, but 
does have more faint sources. In all samples, 
various selections have been made in limiting magnitudes and magnitude
errors, and even redshift in the case of the spectroscopic samples.
A large unknown is the fraction of SDSS extended class sources that are
misclassified point sources, as these potentially add many counts to 
the fainter end of the distributions. In our sample, the extended 
sources are by far the largest group. Richards et al claim that
some 10\% of sources fainter than g$\sim$21 are misclassified point sources.
We find that 17\% or more of them have colours consistent with QSOs. 
Figure 5 shows the distribution of we add this fraction of the MISE to the 
MISP. Overall, we suggest that the source counts are uncertain by a 
factor of order 2 at the faint end.

\newpage
\centerline{\bf References}

Bianchi L. et al, 2007, ApJS, Dec 2007, Astro-ph/0611926

Hutchings J.B., Cherniawsky A., Cutri R.M., Nelson B.O., 2006, AJ, 131, 680
      
Richards G.T. et al, 2004, ApJS, 155, 257

Richards G.T. et al, 2005, MNRAS, 360, 839 (Astro-ph/0504300)

Richards G.T. et al, 2006, Astro-ph/0601434

Trammel G.B. et al, 2006, Astro-ph/0611549

van den Berk D.E. et al, 2005, Astro-ph/0501113

\newpage

\centerline{\bf Captions to Figures}

1. Colours with redshift for SDSS spectroscopic QSOs. The lines are
models using standard QSO SEDs and the SDSS filters. The dashed lines
indicate the colour cuts used in our photometric samples.

2. g magnitude distributions for the spectroscopic samples and photometric
candidates, for the different catalogues as labelled. 

3. Space density of QSOs inferred from the photometric samples, as discussed
in the text.

4. Magnitude SEDs and differences, from SDSS and GALEX filters for the
photometric samples as labelled. The AIS QSOs are overall redder than the MIS. 

5. Space density distributions for our samples, compared with Richards et al
2004. The AIS sample is incomplete at all magnitudes. The difference between
the full and subsample Richards points may show the errors inherent in
samples of 70-80 square degrees.

\newpage

\begin{deluxetable}{lllll}
\tablenum{1}
\tablecaption{GALEX/SDSS candidates and confirmed - database sizes}
\tablehead{\colhead{Property} &\colhead{MIS ext} 
&\colhead{MIS point} &\colhead{AIS ext} &\colhead{AIS point} }
\startdata
Total candidates &29287  &5960   &2633  &3257\cr
Sky area (sq deg) &~~~~~~~~~~~~83 &&~~~~~~~~~~~~363\cr
\hline
Spectra in total &0.2\%  &14\%  &16\%  &47\%\cr
QSO spectra$^a$   &34 &813 &109 &1497\cr
\% QSO spectra &45\% &85\% &25\% &81\%\cr 
Galaxy spectra &42 &8 &317 &21\cr
Star spectra &0 &132 &8 &335\cr
\hline
QSO z $<$ 0.7  &30  &154 &103  &493\cr
QSO z 0.9 to 1.6$^b$ &&412 &&623\cr
QSO z $>$ 1.44$^c$  &&213 &&201\cr
\hline
QSOs, color cuts\cr
g-r full range$^a$    &21  &762   &58  &1438\cr
g-r plus UV-opt$^b$ &12  &382  &38  &508\cr
r-i plus UV-opt$^c$ &21  &235  &57  &180\cr
\hline
Selected candidates\cr
r-i, UV-opt  &(10554)  &1120 &(1575) &230\cr
g-r, UV-opt  &(4945)  &1646  &(1188)  &661\cr

\enddata
\tablenotetext{}{Matching superscripts indicate numbers to be compared.\\ 
Numbers in parentheses are from extended sources with same cuts.}
\end{deluxetable}

\begin{deluxetable}{lcccccc}
\tablenum{2}
\tablecaption{Properties of point source samples}
\tablehead{\colhead{Property} &\colhead{MIS full} 
&\colhead{MIS cand} &\colhead{MIS QSOs} &\colhead{AIS full} &\colhead{AIS cand}
&\colhead{AIS QSOs}\\
&\colhead{all UV} &\colhead{gri cuts} &\colhead{spectra} &\colhead{all UV}
&\colhead{gri cuts} &\colhead{specta} }
\startdata
Mean FUV mag &22.4 &22.8 &21.3 &20.6 &20.3 &20.2\cr
Mean g mag   &20.9 &20.8 &19.2 &19.2 &19.0 &18.8\cr
Mean NUV-r &0.76 &0.95 &0.87 &0.44 &0.80 &0.62\cr
Mean g-r   &0.28 &0.26 &0.16 &0.04 &0.20 &0.12\cr
Mean z meas &&&1.14 &&&0.95\cr

\enddata

\end{deluxetable}

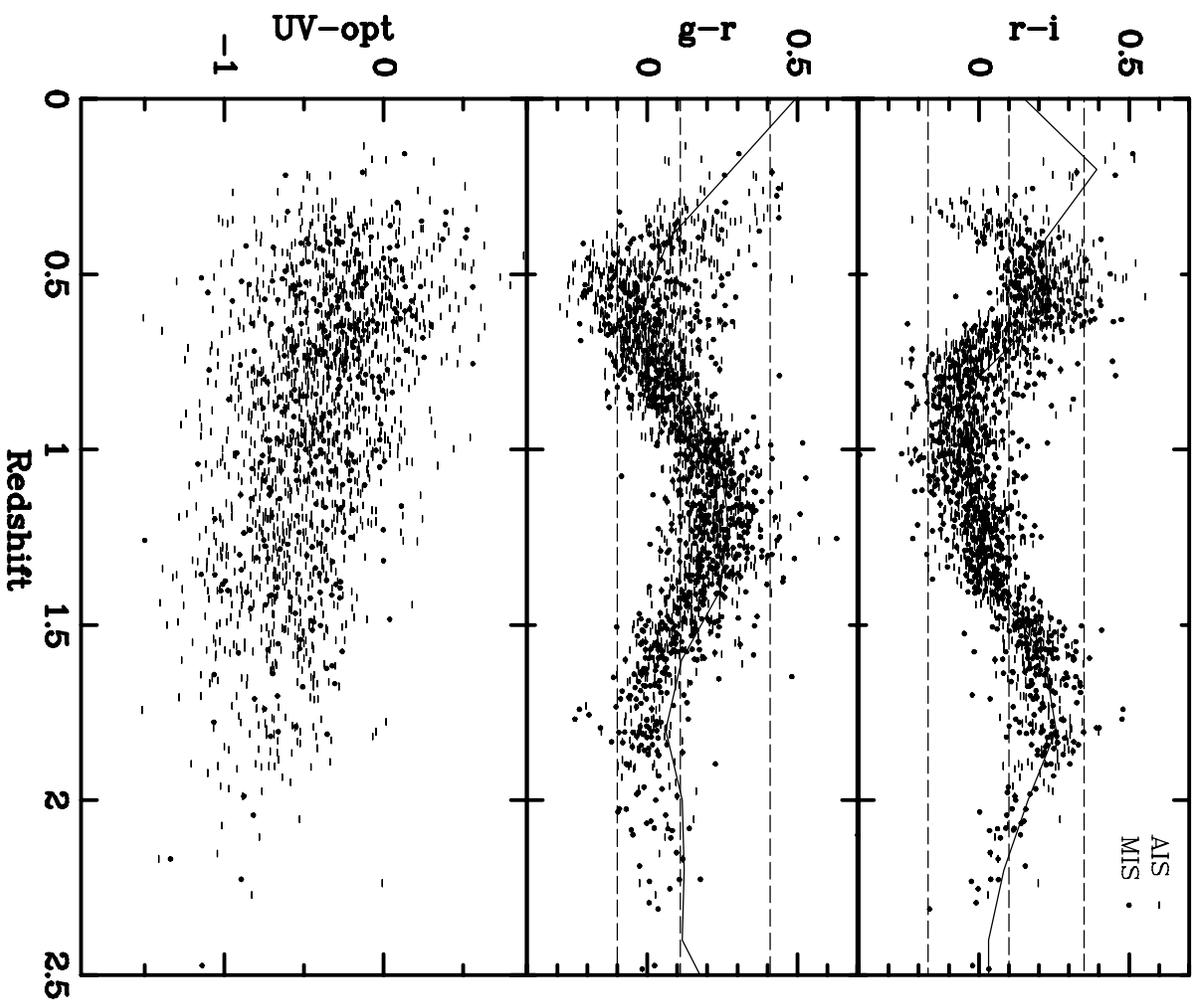
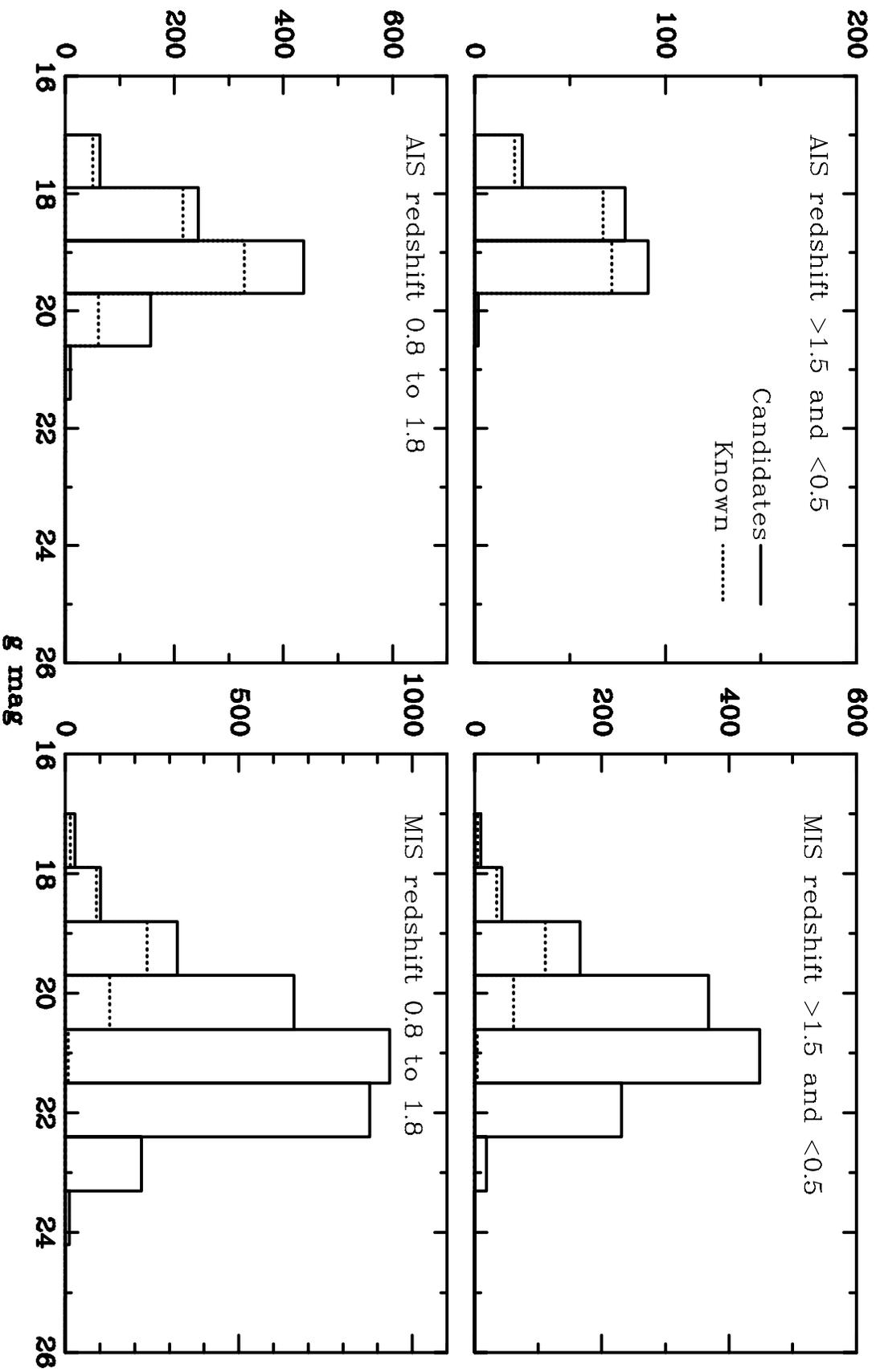
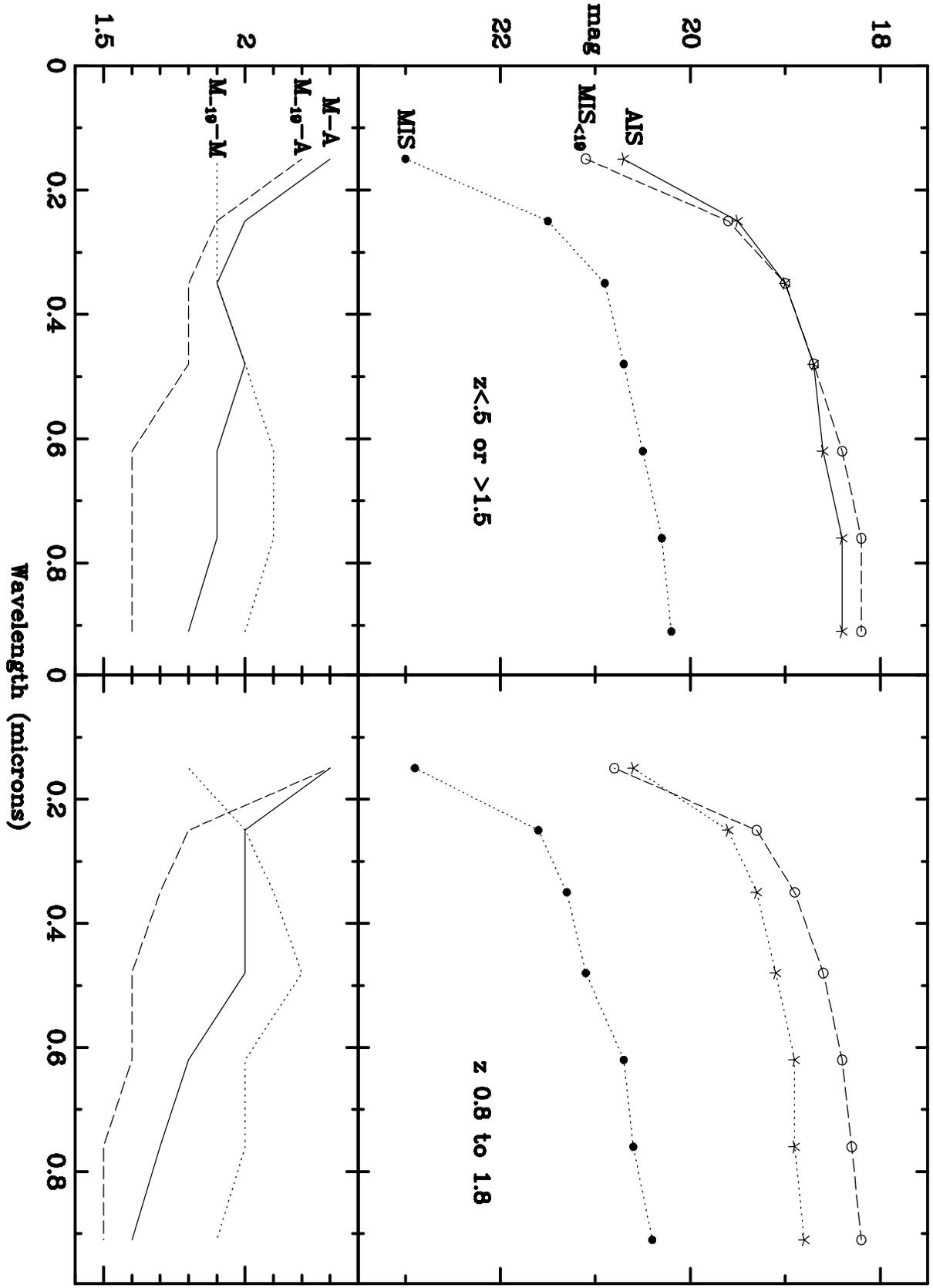
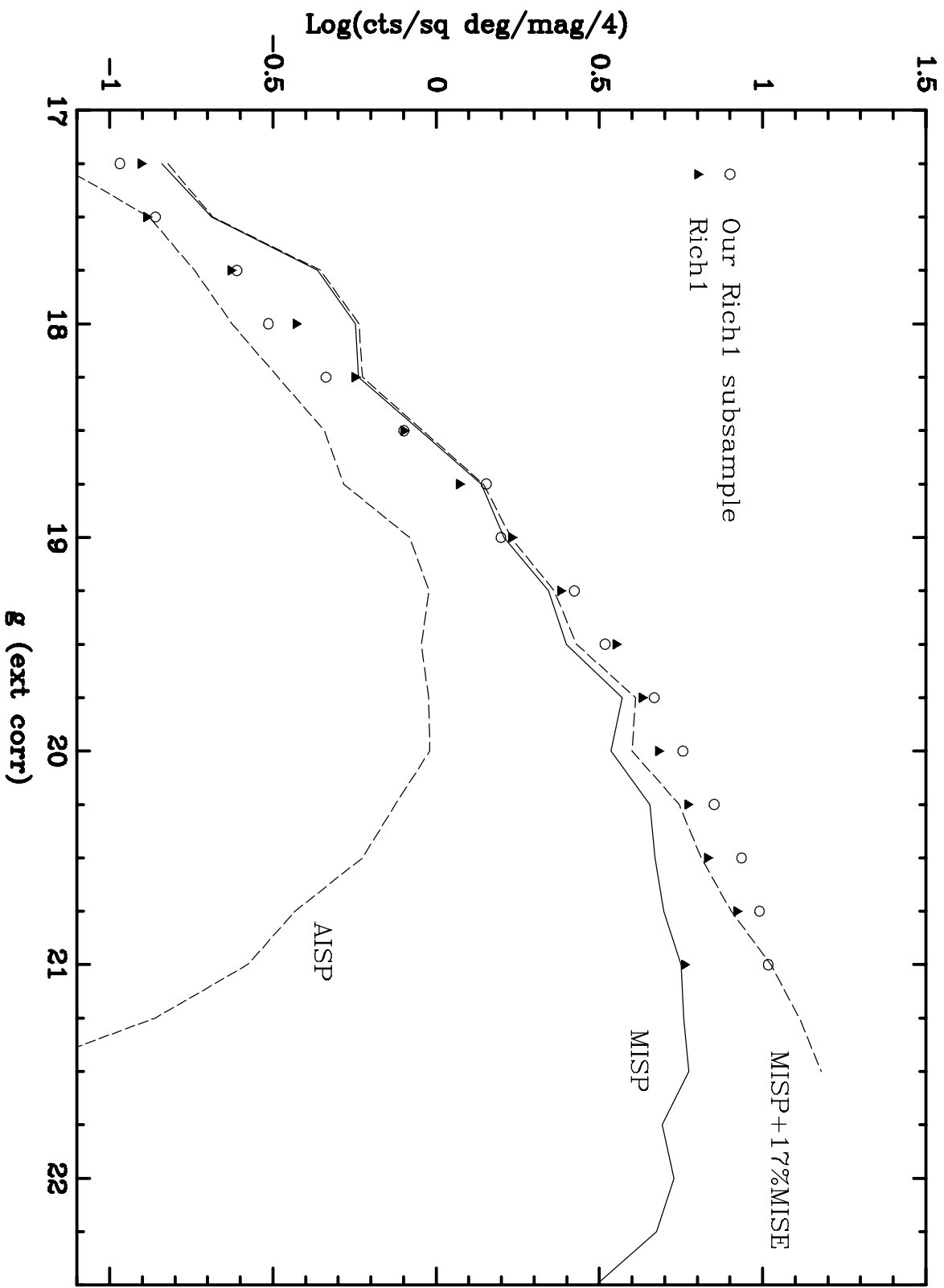
\end{document}